\documentclass[twoside]{ilcws08}
\usepackage[latin1]{inputenc}
\usepackage{graphicx}
\usepackage{subfigure}
\usepackage{wrapfig,rotating}
\usepackage{amssymb,amsmath,array}
\usepackage{units}
\begin{document}
\title{AHCAL Energy Resolution}

%***********************************************************************
% AUTHORS INFORMATION AREA
%***********************************************************************

\author{Katja Seidel$^{1,2}$ for the CALICE collaboration
% DO NOT MODIFY THE FOLLOWING '\vspace' ARGUMENT
\vspace{.3cm}\\
% Addresses and institutions (remove "1- " in case of a single institution)
1- Max-Planck-Institut f\"ur Physik, Munich, Germany\\
2- Excellence Cluster 'Universe', TU M\"unchen, Garching, Germany
}
%%***********************************************************************
% END OF AUTHORS INFORMATION AREA
%***********************************************************************

\maketitle

\begin{abstract}
The CALICE collaboration has constructed highly granular hadronic and electromagnetic calorimeter prototypes to evaluate technologies for the use in detector systems at a future Linear Collider. The hadron calorimeter uses 7608 small scintillator cells individually read out with silicon photomultipliers. This high granularity opens up the possibility for precise three dimensional shower reconstruction and for software compensation techniques to improve the energy resolution of the detector. We discuss the calibration procedure for the analog hadronic calorimeter and present two software compensation methods based on reconstructed clusters, which were developed with simulations and are applied to hadronic test beam data.

\end{abstract}

\section{Introduction}
The CALICE collaboration \cite{calice} has constructed highly granular calorimeter prototypes to evaluate technologies for the use in detector systems at a future Linear Collider \cite{ilc}. The presented results were obtained with pion beams at CERN in 2007 and a calorimeter setup consisting of a silicon-tungsten electromagnetic calorimeter (ECAL) \cite{ecal}, an analog scintillator-steel hadron calorimeter (AHCAL) \cite{ahcal} and a scintillator-steel tail catcher and muon tracker (TCMT) \cite{tcmt}.\\
The ECAL has a total depth of 24 $X_0$ and consists of 30 layers with three different sampling fractions. Each layer has an active area of $\unit[18\times18]{\textrm{cm}^2}$, which is read out by $\unit[1\times1]{\textrm{cm}^2}$ readout pads. The AHCAL consists of 38 \unit[3]{cm} thick stainless steel absorber plates and active areas of scintillator tiles with three different sizes, which are read out with silicon photomultipliers (SiPMs). The tile sizes vary from $\unit[3\times3]{\textrm{cm}^2}$ in the middle to $\unit[12\times12]{\textrm{cm}^2}$ at the outer edges. The last eight layers consist only of tiles with sizes of $\unit[6\times6]{\textrm{cm}^2}$ and $\unit[12\times12]{\textrm{cm}^2}$. In total the AHCAL has a thickness of 5.3 nuclear interaction lengths and uses 7608 scintillator tiles. The TCMT consists of 16 layers of 20 $\unit[100\times5]{\textrm{cm}^2}$ scintillator strips, read out by SiPMs, between steel absorber plates. The first 8 absorber layers have a thickness of 19\,mm the last 8 layers have \unit[102]{mm} thick plates. The orientation of the scintillator strips alternates between vertical and horizontal in adjacent layers.
\\
In the following, preliminary results of the CALICE collaboration are presented on the calibration of the analog hadron calorimeter and on new studies of software compensation techniques to improve the energy resolution of hadron showers in the AHCAL and the TCMT.

\section{Calibration of the AHCAL}
\label{sec:calibration}
The main calibration steps of the CALICE analog hadron calorimeter are motivated by the SiPM characteristics and consist of a signal saturation correction, a calibration of the SiPM gain using LED light, a MIP-calibration using muon data and a temperature correction of the SiPM amplitude.\\
The finite number of pixels of the SiPM limits the maximum number of detectable photons. This results in a non-linear dependency between a light signal which is coupled to the SiPM and the SiPM response. Therefore the SiPM saturates for large light signals. The saturation of each SiPM which is  installed in the calorimeter was measured on a test bench. These saturation measurements allow to correct the SiPM signal amplitude. \\
The gain of each SiPM is measured using an LED system. Low-intensity LED light is coupled in the scintillator tile. The LEDs are placed outside the calorimeter and the light is connected via fibers to the scintillator tiles. A signal spectrum of single photon peaks is measured and the distance between two single neighboring single photon peaks gives the gain of the device.\\
To have an equal cell response to a minimum ionizing particle (MIP), the energy deposition of high energy muons is measured. The whole detector was illuminated with high energy muons and the measured signals for each calorimeter cell were equalized to have the MPV at the same position. The calibrated energy scale is therefore given in units of the energy deposition of  one minimum ionizing particle. \\
The signal of a SiPM is very temperature sensitive. The analog hadron calorimeter was therefore equipped with temperature sensors, which measured the temperature during run time. The measured amplitudes are corrected for the known temperature dependence of the SiPM response.
\begin{figure}[!htb]
	\subfigure{\includegraphics[width=0.3\textwidth]{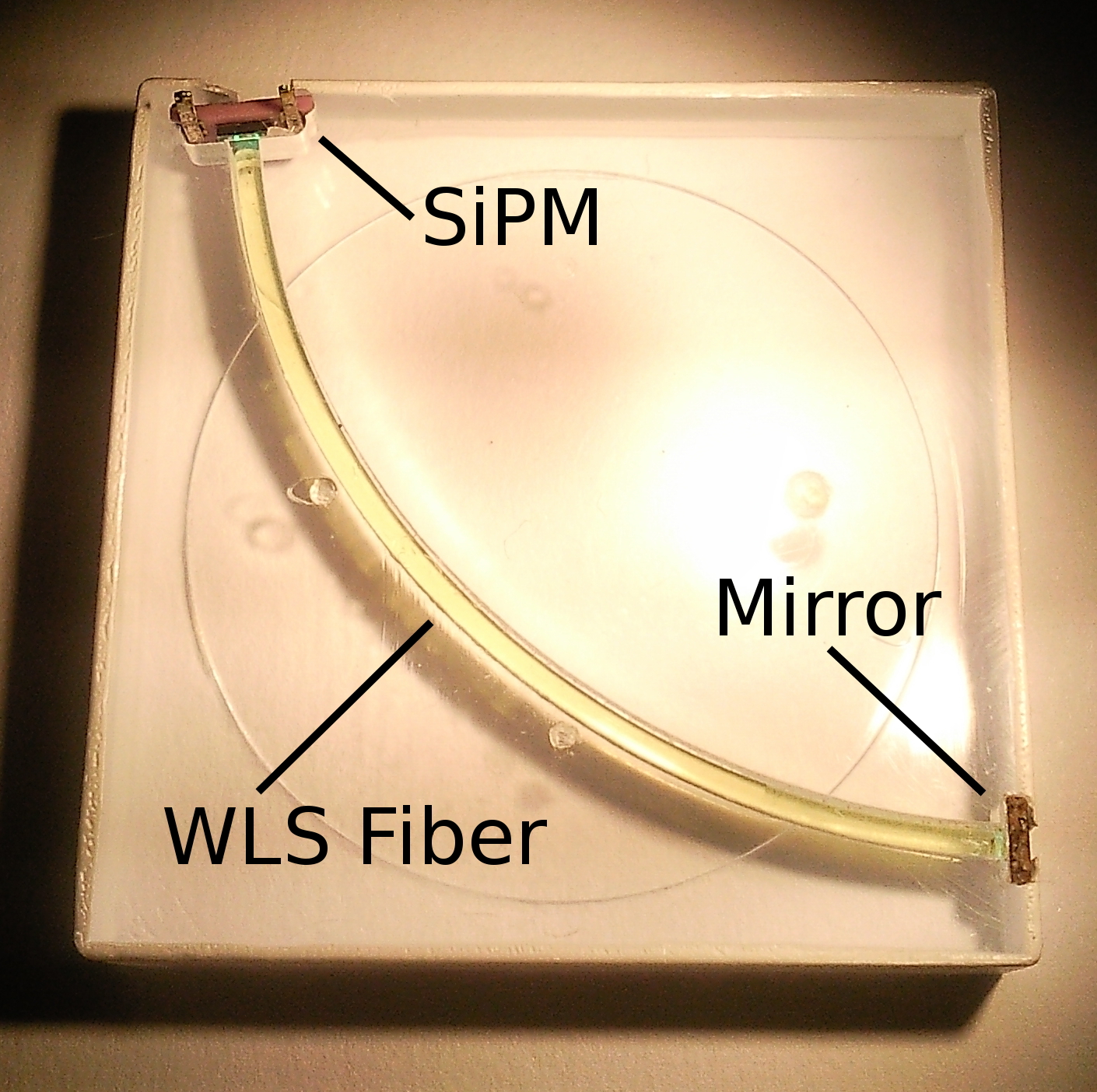}}\hfill
	\subfigure{\includegraphics[width=0.45\textwidth]{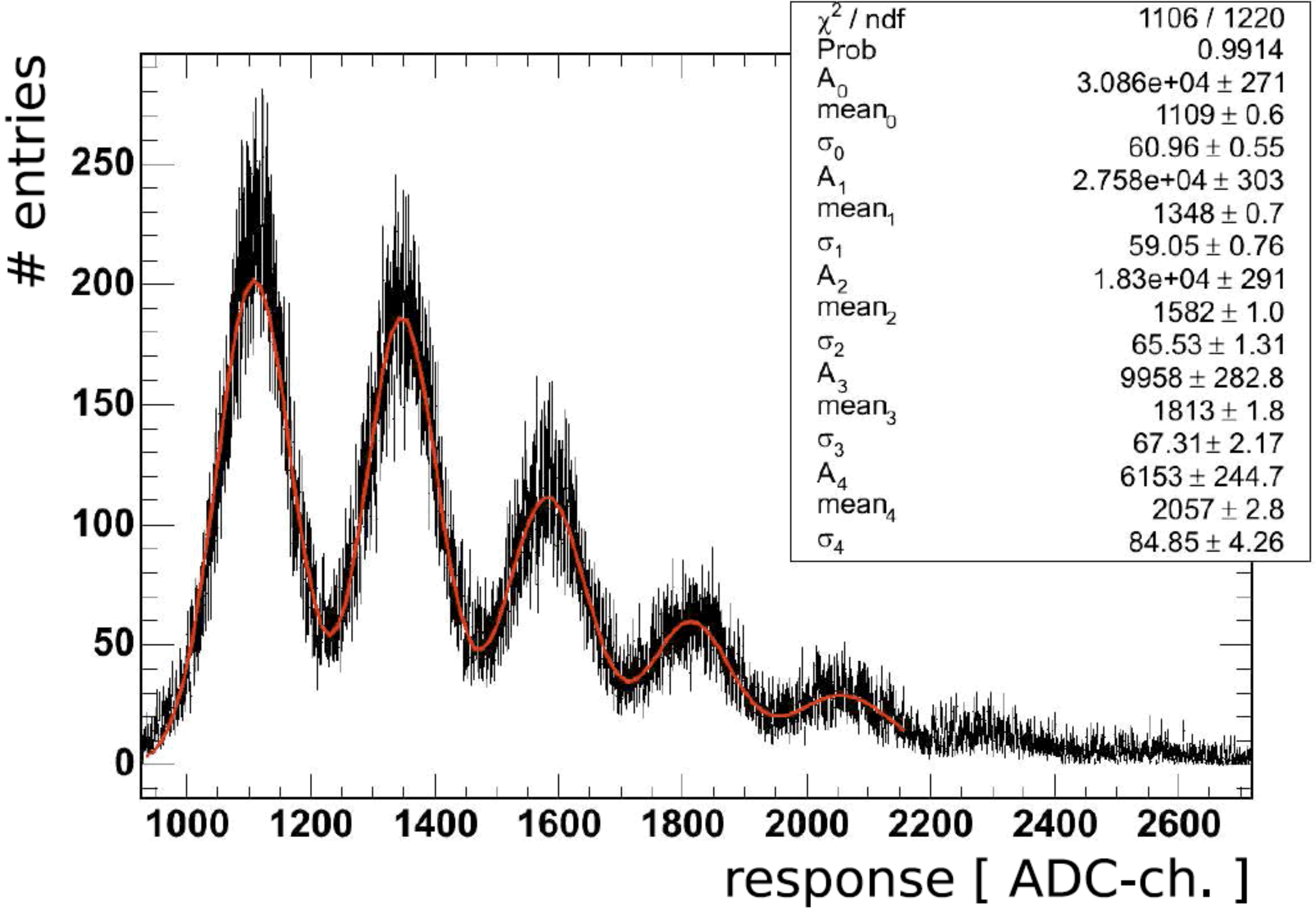}}
	\caption{Left: Scintillator tile of the AHCAL with wavelength shifting fiber and SiPM. Right: Single photon spectrum of one SiPM to measure the gain of the SiPM.}
	\label{fig:calib}
\end{figure}

\section{Software Compensation Techniques}
Software compensation techniques can be used to improve the single particle energy resolution of the CALICE calorimeters. In non-compensation calorimeters like the CALICE detectors, the measured signal of electrons is typically larger than of hadrons. In order to improve the energy reconstruction of hadron data, a weighting technique was developed for showers in the AHCAL and the TCMT. This technique exploits that electromagnetic showers have a higher energy density. The overall density of the reconstructed cluster is taken therefore as a measure for the electromagnetic content. Showers with a higher electromagnetic content get a lower weight in the energy reconstruction than showers with a small electromagnetic fraction. The second presented technique uses a neural network to reconstruct the cluster energy from cluster variables which are used as an input for the neural network.

\subsection{Cluster Density Weighting Technique}
\begin{figure}[!htb]
	\subfigure{\includegraphics[width=0.5\textwidth]{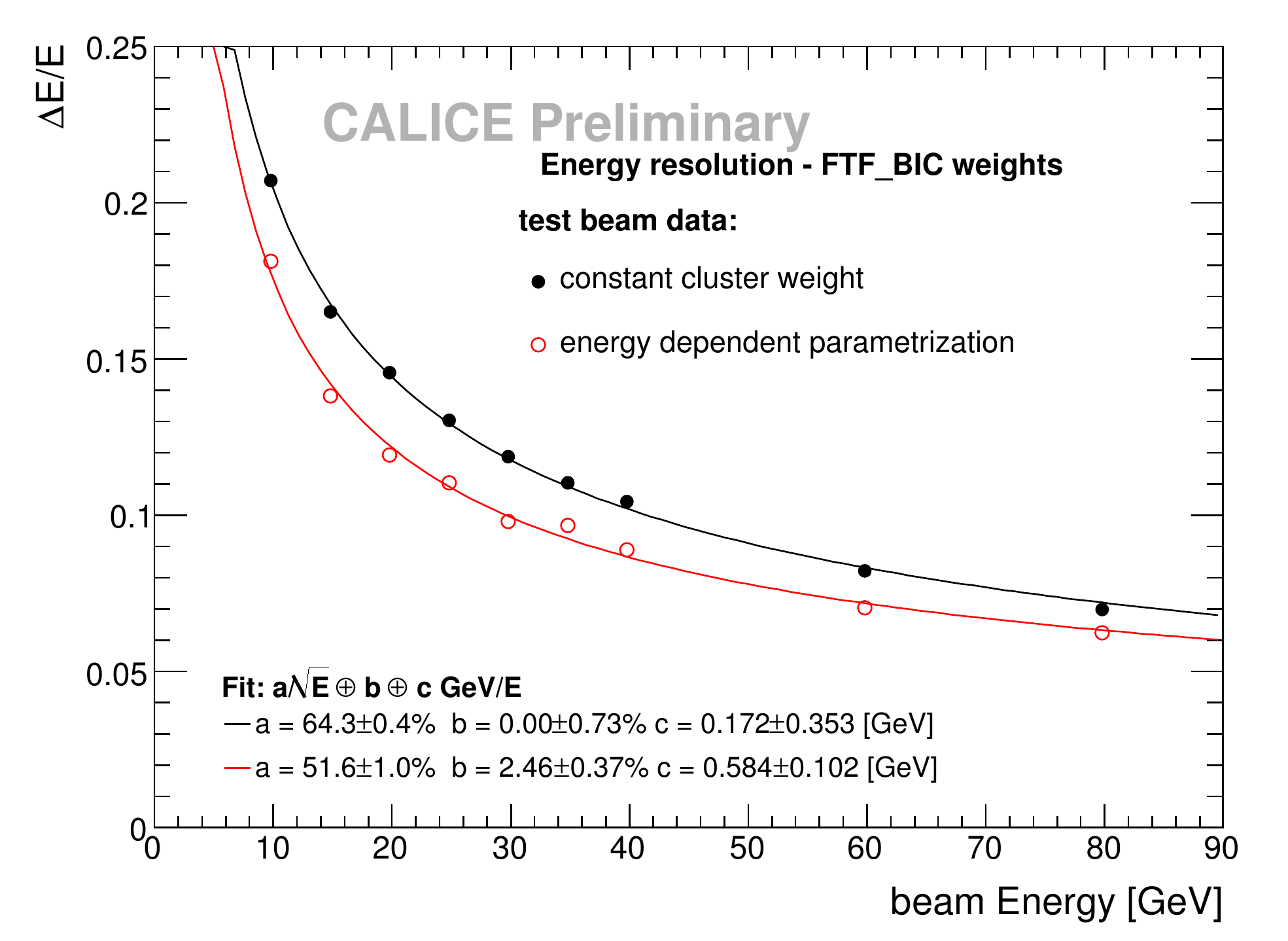}}\hfill
	\subfigure{\includegraphics[width=0.5\textwidth]{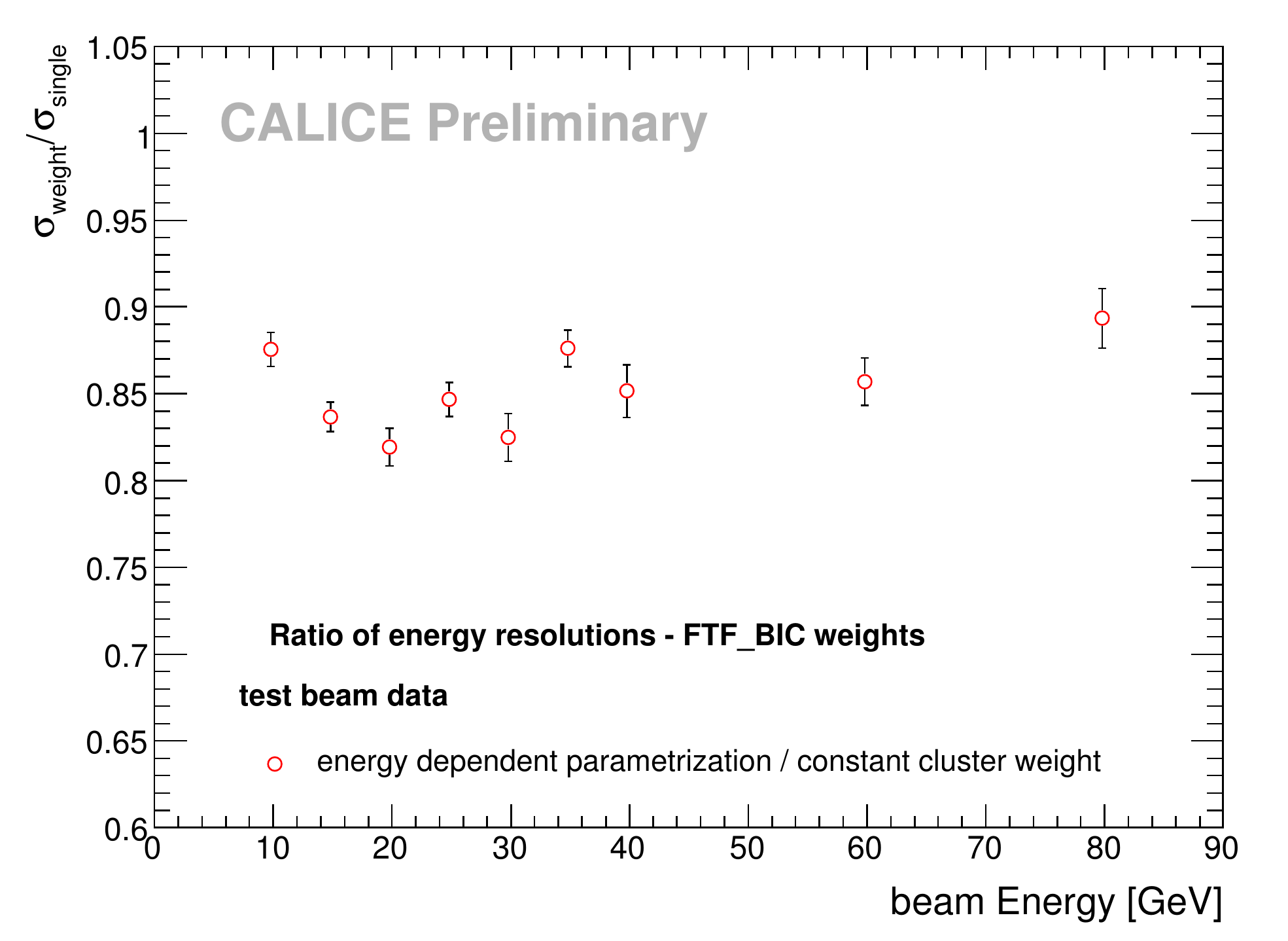}}
	\caption{Left: Energy resolution of test beam data before and after the software compensation using a cluster energy density weighting technique applied. The weight functions were extracted with Monte Carlo data of the hadronic model FTF\_BIC. Right: Ratio of energy resolution before and after the software compensation.}
	\label{fig:resolutionW}
\end{figure}

To reconstruct the energy of a particle shower in the calorimeter, a conversion factor is used to transform from the calibration scale (MIP) to the energy in GeV. The developed software compensation method uses non constant cluster weight factors. The weight factor for each cluster depends on the cluster energy density (cluster energy/cluster volume) as a measure of the elctromagnetic content of the shower. Clusters with a higher energy density get assigned a lower weight/conversion factor to calculate the cluster energy on the GeV scale. The weights have to be energy dependent, therefore the appropriate weight is calculated using the energy density and the reconstructed energy of the cluster (in MIPs) as an input.  Weights are determined via functions and the parameters of these functions were determined with Monte Carlo simulations using GEANT4 with the QGSP\_BERT and FTF\_BIC physics lists. The parameters were choosen in order to achieve the best reconstructed energy of the Monte Carlo data. To reconstruct the cluster energy of test beam data the weight function extracted from Monte Carlo is used. The results of the software compensation using weights extracted from FTF\_BIC is shown in Figure \ref{fig:resolutionW} and \ref{fig:linearity}.
The energy resolution improved by approximately \unit[13]{\%} compared to the constant cluster weight method and the reconstructed energy matches within \unit[3]{\%} with the beam energy, as shown in Figure \ref{fig:linearity} left.

%%%%%%%%%%%%%%%%%%%%%%%%%%%%%%%%%%%%%%%%%%%%%%%%%%%%%%%%%%%%%%%%%
%% NN
%%%%%%%%%%%%%%%%%%%%%%%%%%%%%%%%%%%%%%%%%%%%%%%%%%%%%%%%%%%%%%%%%
\subsection{Neural Network Technique}
\begin{figure}[!htb]
	\subfigure{\includegraphics[width=0.5\textwidth]{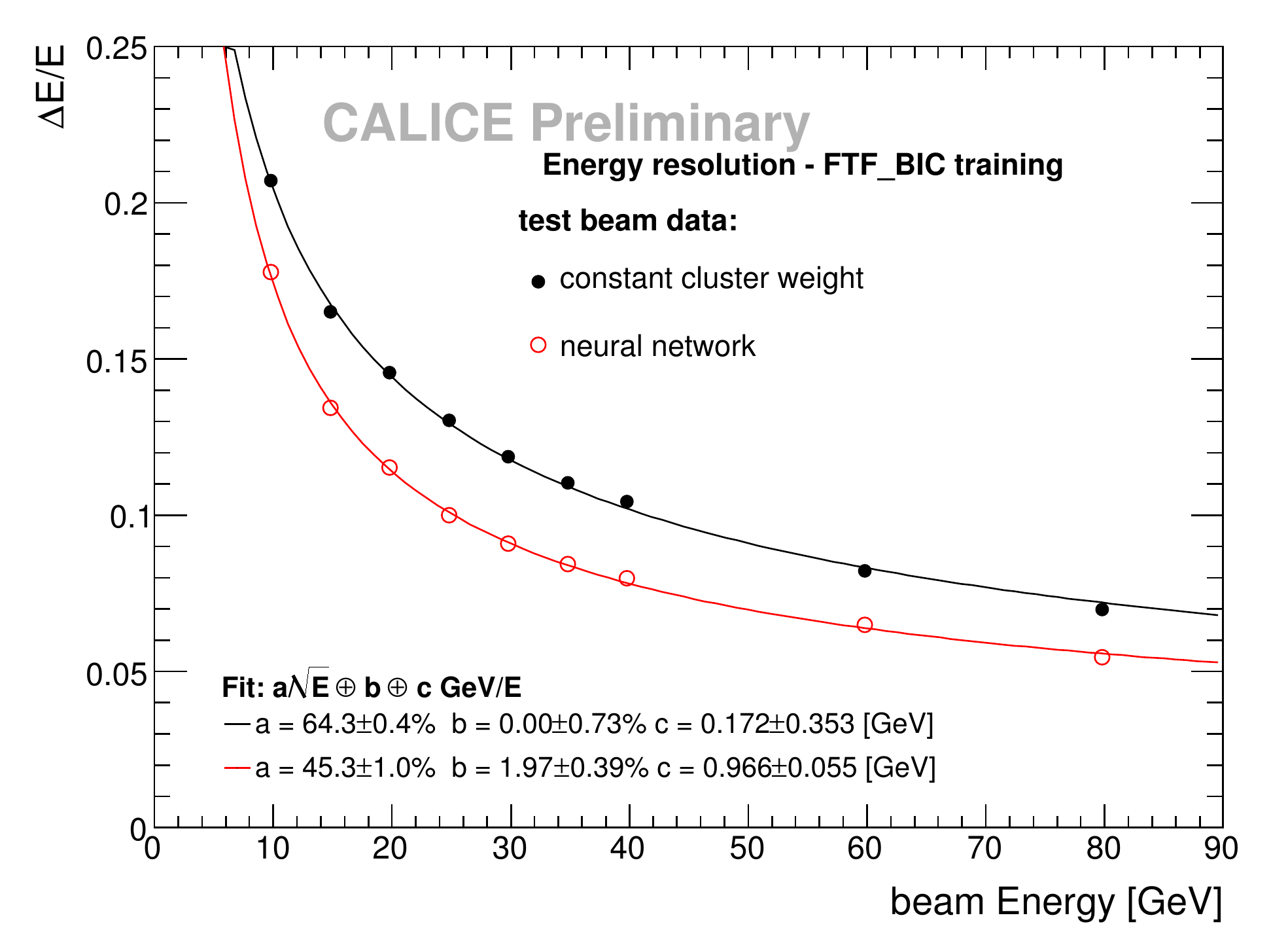}}\hfill
	\subfigure{\includegraphics[width=0.5\textwidth]{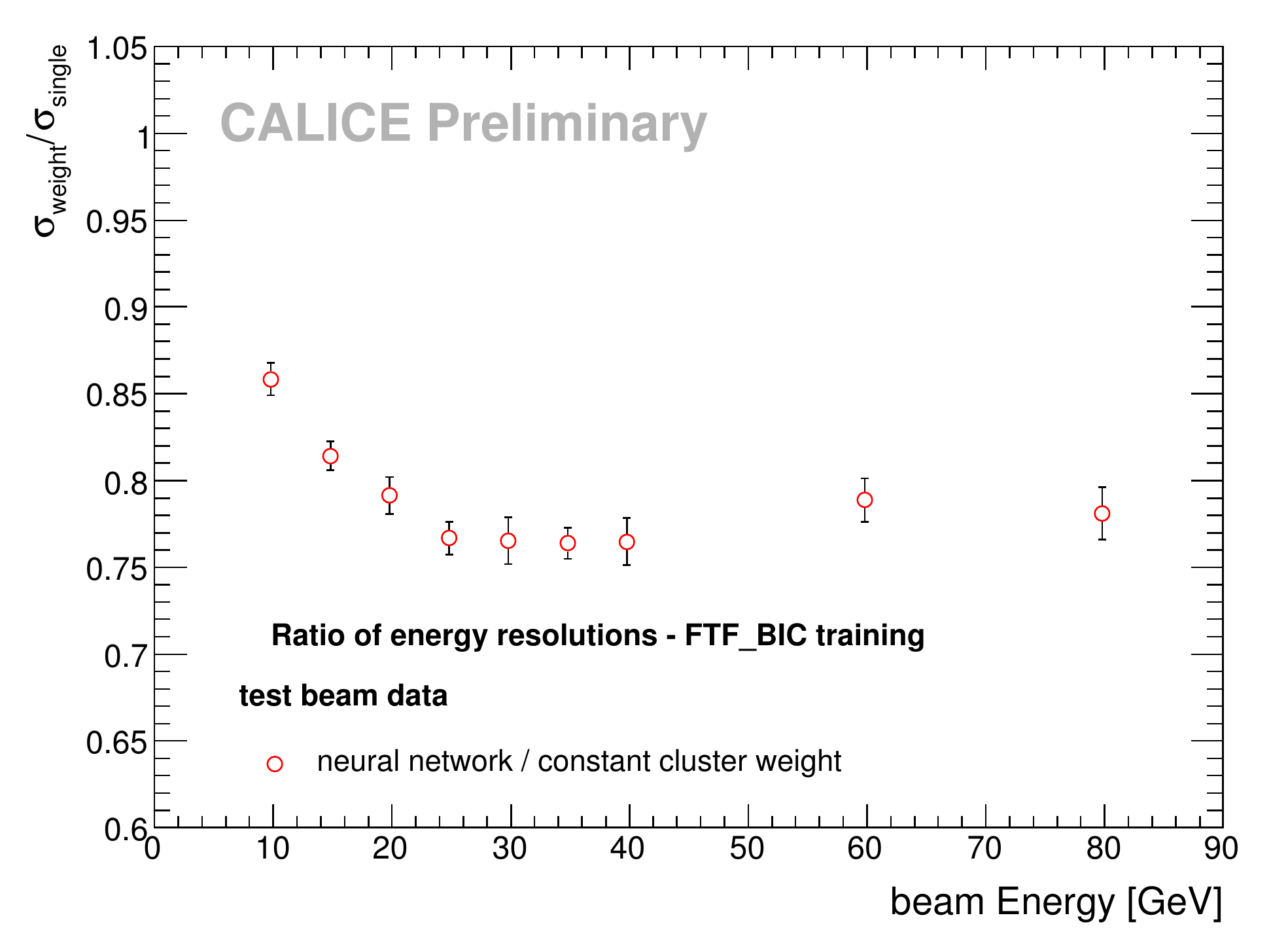}}
	\caption{Left: Energy resolution of test beam data before and after the software compensation using a neural network. The neural network was trained with Monte Carlo data of the hadronic model FTF\_BIC. Right: Ratio of the energy resolutions before and after software compensation.}
	\label{fig:resolutionNN}
\end{figure}

\begin{figure}
	\subfigure{\includegraphics[width=0.45\textwidth]{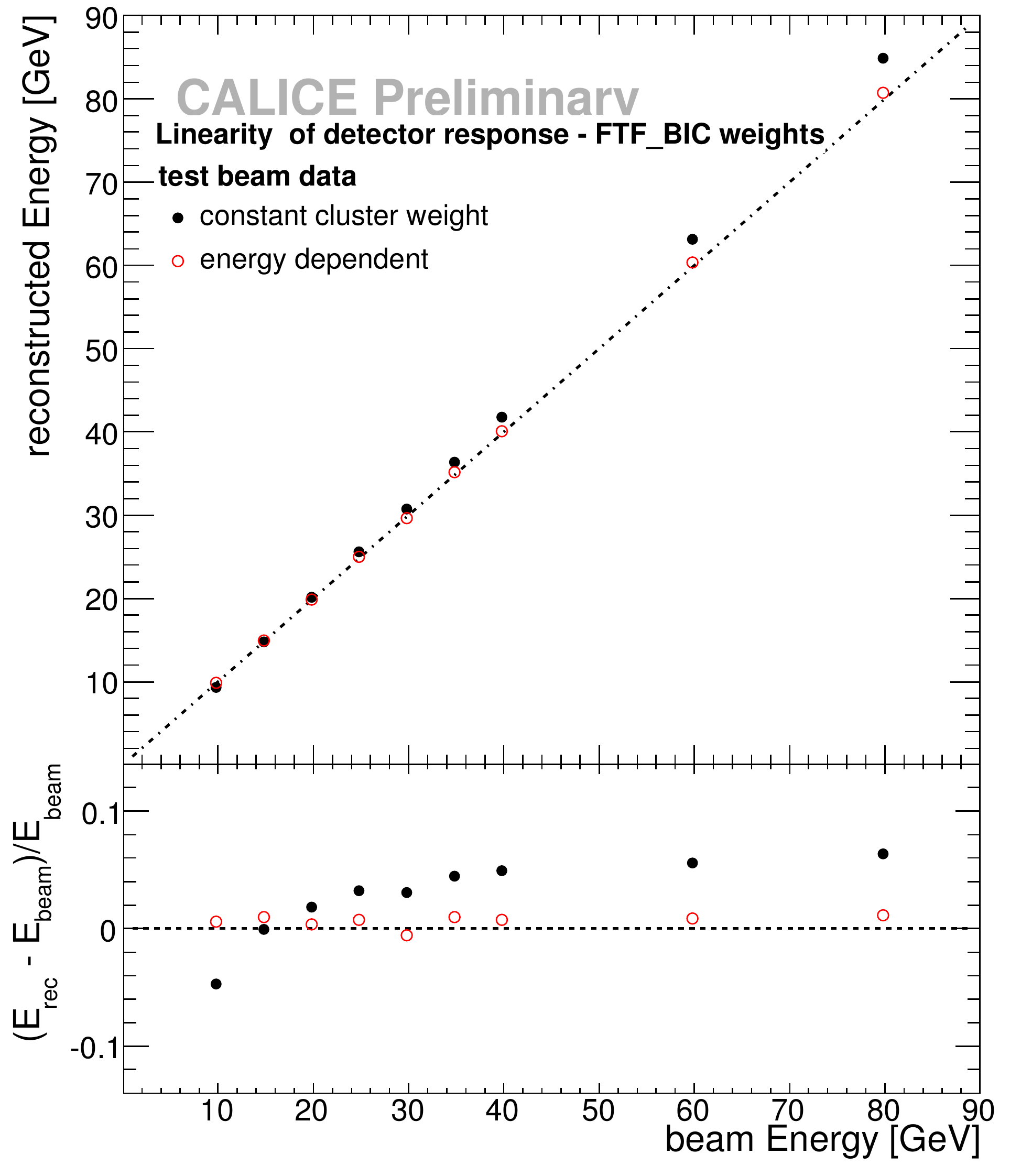}}\hfill
	\subfigure{\includegraphics[width=0.45\textwidth]{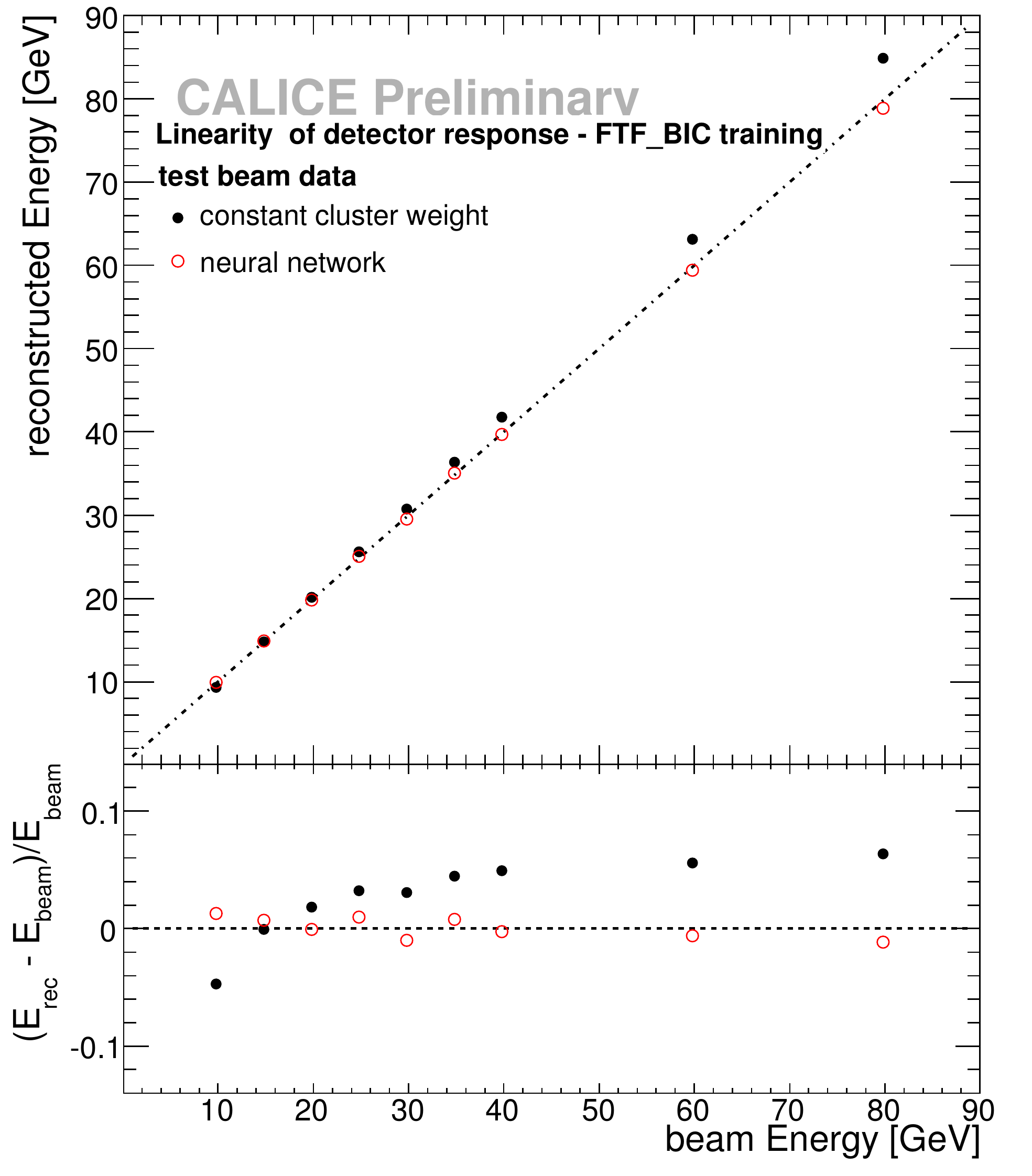}	}
	\caption{Linearity of the detector response of test beam data before and after the software compensation Left: Use of a cluster energy density weighting technique. The weight functions were extracted with Monte Carlo data of the hadronic model FTF\_BIC. Right: Use of a neural network. The neural network was trained with Monte Carlo data of the hadronic model FTF\_BIC. }
	\label{fig:linearity}
\end{figure}

The other studied software compensation technique uses a neural network to reconstruct the energy of the hadronic particle. The neural network was trained with the same Monte Carlo set as discussed above. The simulated data set had to have a nearly continous energy range to avoid that the neural network is trained for certain energy steps. Six input  variables (cluster energy, cluster length, cluster volume, cluster width, cluster energy in the last five AHCAL layers and cluster energy in the tail catcher) were used in the neural network. During the training phase of the regression method of the neural network the output variable, the reconstructed energy in GeV, was trained to match the target variable, the beam energy. The target variable is not used during the application of the trained neural network. The energy reconstruction of test beam data using a neural network which was trained with Monte Carlo data of the hadronic model FTF\_BIC is shown on the right side of Figure \ref{fig:linearity}. The linearity is better than \unit[2]{\%} over the studied energy range.
Also the energy resolution improves significantly compared to constant cluster weight energy reconstruction. The gain in energy resolution, shown by the ratio of the energy resolution with and without the neural network applied in Figure \ref{fig:resolutionNN}, is approximately \unit[23]{\%}.

\subsection{Energy resolution of the complete CALICE calorimeter setup}
The cluster based software compensation techniques can be compared with a weighting technique based on the single cell energy density. In this method a different weight/conversion factor for every hit energy in the overall energy sum is used to calculate the energy in GeV. The weight of the single hit depends on the cell density and is energy dependent. With this technique an improvement in the energy resolution of \unit[20]{\%} was achieved and the linearity of the detector response was significantly improved for the whole CALICE calorimeter setup (ECAL + HCAL + TCMT) \cite{frank}.

\section{Summary}
Non-compensation calorimeters with a high granularity have been constructed and tested by the CALICE collaboration. The calibration steps of the analog hadron calorimeter were presented. The high granularity allows to precisely determine shower variables which are used in a neural network and a weighting technique of software compensation methods. These techniques were calibrated using Monte Carlo data and were used to calculate the cluster energy of hadron test beam data in the AHCAL and the TCMT. The energy resolution and the linearity of the detector response were significantly improved using these techniques. The neural network based technique, which uses several shower parameters, shows significantly better performance than the single weighting technique based on the cluster energy density alone.

\end{document}